\documentclass[11pt]{article}
\begin{document}
\begin{center}
\today \\
\begin{Large} {\bf
Periodicity, Thermal Effects, and Vacuum Force: Rotation  in Random Classical
Zero-Point Radiation.}
\end{Large}
\vspace{10mm}\\
\begin{it}
{Yefim S. Levin }\\
Department of Electrical and Computer Engineering, Boston University,
Boston, MA, 02215   \\
\end{it}
\end{center}
\begin{abstract}
\indent Thermal effects of acceleration through a vacuum have been investigated
in the past from different perspectives, with both quantum and classical methods. However,  the existence  of the thermal effects  associated with rotation in a flat vacuum  requires a deeper analysis. In this work
we show that for a detector rotating in a random classical zero-point electromagnetic or massless scalar radiation at zero temperature such thermal effects exist.  Analysis and calculations are carried out in terms of correlation functions of random classical electromagnetic or massless scalar field in the rotating reference system. This system is constructed as an infinite set of Frenet-Seret tetrads $\mu_{\tau}$ defined so that the detector is at rest in a tetrad at each proper time $\tau$. Particularly,  (1) correlation functions, more exactly their frequency spectrum ,
 contain the Planck thermal factor
$1/(\exp (\hbar \omega /k_B T_{rot})-1)$, and (2) the energy density the rotating detector observes is proportional to the sum of  energy densities of
Planck's spectrum at the temperature $T_{rot} = \frac{\hbar \Omega}{2 \pi k_B} $ and zero-point radiation. The
proportionality factor is $\frac{2}{3}(4 \gamma^2 -1 )$ for an  electromagnetic
field and $\frac{2}{9}(4 \gamma^2 -1 )$ for a massless scalar field, where $\gamma = (1 - (\frac{\Omega r}{c})^2)^{-1/2}$, and $ r $ is a detector rotation radius.
The origin of these thermal effects is  the periodicity of the correlation functions and their discrete spectrum, both following rotation with angular velocity $\Omega$. The correlation functions without periodicity properties do not display thermal features. The thermal energy can also be interpreted as a source of a force, $f_{vac}$, applied to the rotating detector from the vacuum field, $``vacuum \; force"$.
The $f_{vac}$ depends on the size of neither the charge nor the mass,
like the force in the Casimir model for a charged particle, but, contrary to the last one, it  is directed to the center of the circular orbit.
The  $f_{vac}$ infinitely grows by magnitude when $r \rightarrow \; r_0 = c/ \Omega$. Therefore the radius of circular orbits with a fixed $\Omega$ is bounded. The orbits with a radius greater than $r_0$ do not
exist simply because the returning vacuum force becomes infinite. On the uttermost orbit with the radius
$r_0$, a linear velocity of the rotating particle would have become $c$.  The  $f_{vac}$ becomes very small and proportional to $r$ when $r$ is small,
$r \ll c/\Omega$. Such vacuum force dependance on radius, at large and small $r$, can be associated respectively with so called confinement and asymptotic freedom, known in quantum chromodynamics, and provide a new explanation for them.
\end{abstract}
\section{Introduction.}
\indent This work is focused on thermal effects hypothetically associated with rotation  through a vacuum of a massless scalar or electromagnetic field in a flat, Minkowski, space, and performed in a classical approach.  \\
 \indent Investigations of rotation  are  mostly
based on  the ideas developed for a linear acceleration through a vacuum
\cite{davis1996} - \cite{ylevin2009}.
For example, in \cite{davis1996}, the authors write: ``... in the Rindler case, a set
of uniformly accelerated particle detectors ... will give zero response
in the Rindler vacuum state, and will give a consistent
thermal response to the Minkowski vacuum state."
And later on: ``We might therefore expect a set of rotating detectors to
similarly reveal the state of a rotating vacuum field". This program
for the rotation case was used, for example, in \cite{davis1982} and
\cite{pfautsch1981}  and after that in \cite{lorency2000}.  Below we discuss
some results of \cite{lorency2000}. \\
\indent A rotating 4-space,  with the Trocheries-Takeno (T) coordinates and a non-static non-diagonal metrics, with the associated  quantum Fock space (referred below as T-F space) of a massless scalar field are considered in \cite{lorency2000}, along with
Minkowski (M) space and its associated  Fock (M-F) quantum space. \\
\indent The T-coordinates
$(t,r, \theta, z)$ are
connected with M-coordinates $(\tilde{t}, \tilde{r},
\tilde{\theta}, \tilde{z})$  as:
 \begin{eqnarray}
 \label{eq:TT_Coord}
 t=\tilde{t} \cosh\Omega \tilde{r} - \tilde{r} \tilde{\theta}\sinh\Omega\tilde{r},
 \nonumber \\
 r=\tilde{r}, \nonumber \\
 \theta= \tilde{\theta}\cosh \Omega \tilde{r} -\frac{\tilde{t}}{\tilde{r}}
 \sinh \Omega \tilde{r}, \nonumber \\
 z=\tilde{z}.
\end{eqnarray}
The main motivation to use T-coordinates is that a particle at  rest  in T-space,
with  constant values of $( r, \theta, z ) $,
has a velocity $v(\tilde{r})= \tanh(\Omega \tilde{r})$ in the M-space, which is less than the  speed of light
for any $\tilde{r}$.  \\
\indent The ``rotating vacuum" of a massless scalar
field in the T-F space is not the Minkowski vacuum, $|0_M \rangle \neq |0_T \rangle$,  because
the Bogolubov  coefficients of the transformation between creation-annihilation operators $( b^{+}_M, b_M)$ and
 $( a^{+}_T,a_T)$ of M-F and T-F quantum spaces respectively  are not equal to zero. \\
\indent  Based on this fact, the response function,  $R(E)$,
of a rotating Unruh - De Witt detector  in a massless scalar field  is obtained in \cite{lorency2000}. It
describes probability of excitation of the detector with energy $E$ per unit proper time. The authors
consider the $R(E)$ for  three different situations depending on the motion of the detector and the state in which the quantized field is
prepared. 1.  The  response function,  referred  to  as $R^{(r)}_M(E, R_0)$ \cite{lorency2000},  for the field in the Minkowski vacuum state, $| 0_M \rangle$, and at the detector rotating (r)
in Minkowski space on a circular orbit with radius $R_0$.
 2. The response function, $R^{(i)}_T(E, R_0)$, for the field in the rotating
vacuum state, $|0_T \rangle$,  of T-F quantum space, and the
inertial (i) detector, non rotating,  at the distance $R_0$ from the field rotation center in M-space.
3. The response function, $R^{(r)}_T(E, R_0)$, for the field in  the rotating vacuum state, $|0_T \rangle$, of T-F quantum space and the rotating (r) detector in Minkowski space on an orbit with radius $R_0$.\\
\indent The results obtained for the first two scenarios look self-consistent and meet
the expectations based on
the experience gained from the Rindler case of a uniformly accelerated detector, at least in part.
They still do not reveal Planck's thermal properties of a vacuum associated with rotation. \\
\indent The third situation is less clear. This is what authors  say about
it \cite{lorency2000}: ``...we once
again arrive at the same confrontation between canonical quantum field  theory and the detector formalism, which was
settled by Letaw and Pfautsch and Padmanabhan: how is it possible for the
orbiting detector to be excited in the rotating vacuum". The non-null excitation rate in the
third scenario, the authors say in \cite{lorency2000},  can be attributed  to two independent origins:
1. non-staticity of the Trocheries- Takeno metric, and 2. to the Unruh - De Witt detector model adapted in \cite{lorency2000}. The Glauber model detector would not be
excited in this situation. \\
\indent The authors in \cite{lorency2000} give this problem the following explanation and
solution: ``Because the rotating vacuum excites even a rotating detector, we consider this
as a noise which will be measured by any other state of motion
of the detector." And: `` This amounts to saying that the inertial detector will also measure this noise, and we normalize the rate in this situation by subtracting from it the value
of $R^{(r)}_T(E, R_0)$, resulting in a normalized excitation rate for the inertial detector in
interaction with the field in the rotating vacuum." So, instead of $R^{(i)}_T (E,R_0)$,
they
use $\tilde{R}^{(i)}_T(E, R_0) = R^{(i)}_T (E, R_0) - R^{(r)}_T(E, R_0)$. \\
\indent But, even with this correction, there is one more problem associated with rotating vacuum which is  not addressed
in \cite{lorency2000}. In the Minkowski space, none of the points of the rotating
system considered as a ``rotating vacuum" has an angular velocity
$\Omega$. Indeed, the angular velocity $\Omega_M$, in the Minkowski space, of a point
with fixed spatial T coordinates
$(r = R_0,\theta= \theta_0, z=z_0)$ is
\begin{eqnarray}
\Omega_M =\frac{d \tilde{\theta}}{d \tilde{t}} = \frac{1}{R_0} \tanh \Omega R_0 \neq
\Omega.
\end{eqnarray}
So  in Trocheries - Takeno coordinates formalism \cite{lorency2000}, $\Omega$ is just a
parameter without a clear physical sense, and the concept of ``rotating vacuum"  is ambiguous.  \\
\indent In this work we do not use the concept of  ``rotating vacuum". The word ``rotation" is associated with a detector moving on a circle only. Our approach to the problem is based on the concept of ``measurements" made by a point-like detector,
rotating  in a scalar or electromagnetic vacuum.  Bernard suggested in \cite{bernard1986}
``to represent  measurements by an observable, without describing the detection process," with a
``transformation law which tells us how this observable is modified when the same detector
is forced to move along some other world line".
This should be applicable to both quantum and classical theory.
Nevertheless, analysis of local measurements in terms of local observables  only,  without any
references to a detector features, turns out to have some restrictions. Indeed, the character of the
motion of a detector implies some detector features and therefore determines a detection process. For example, in the frame of special relativity theory, a rotating  detector should have a charge to be rotated and held on a circle. Therefore it  behaves like a rotating oscillator and should be selective to frequencies. We will show
 that the
 angular velocity of the observer is a key parameter to describe the thermal properties of the
 rotation in random zero-point classical radiation.\\
\indent Regarding the transformation law of an observable, mentioned above \cite{bernard1986}, to represent a measurement, both quantum and classical, the simplest assumption is  that
the observable is
 an invariant for all possible world lines and coordinate systems. Mathematically an observable with
 such properties can be  described in a tetrad formalism, because  tetrad components of
vectors and tensors are invariants  with respect to coordinate transformations
 \cite{ moller} - \cite{ misner73}\\
\indent A similar approach has been used in \cite{boyer1980} for a uniformly accelerated  observer, even though the tetrad formalism was not used explicitly. Inertial systems, local in terms of time
and each defined at an  observer proper time, were used in \cite{boyer1980}. The tetrad formalism was used in
\cite{cole1986} to describe interacton between two  uniformly accelerated oscillators in a vacuum, located in a plane
perpendicular to the motion direction. \\
\indent In this work, the measurements made by the rotating detector are described in the rotating reference system consisting of an infinite number of
instantaneous inertial reference frames and mathematically defined
as tetrads at each moment of the detector proper time. Along with such a reference system, the two-point correlation functions
of the electromagnetic and scalar massless field and  energy
density of these fields are defined and analyzed for
zero-point radiation.\\
\indent The article is organized as follows. \\
 Section \ref{sec-CorFunForEelctromagneticFieldAtRot} is dedicated to a detector motion through a random classical zero-point electromagnetic radiation.
Subsection \ref{sec:TetradsMeasuments}, Appendixes
\ref{sec:Fermi-Walker} and \ref{sec-generalExpressions}:
The expressions for the components of the electromagnetic field measured at a
Frennet-Seret tetrad are found in terms of the field components in the laboratory inertial coordinate system,  and  correlation functions of the electromagnetic field at a rotating detector are constructed.
Subsection \ref{sec-IntegralExpressions}: The correlation function calculation scheme is described.
Subsection \ref{sec-Final} and Appendix \ref{sec-IntegralCalculationFinal}: The final expressions of the correlation functions in terms of
elementary functions are given. These correlation functions turned out not to display any thermal features.
Subsection \ref{sec:PeriodicityOfCF}: An assumption about the existence of the periodicity of the correlation functions and a discrete spectrum associated with it is discussed and justified. To the best of our knowledge, this idea has not been discussed in the literature yet.
Subsection \ref{sec:CFDiscrete}, Appendix \ref{sec-ModifiedExpressionRandom}: New correlation functions with the discrete spectrum are constructed.
Subsection \ref{sec:Abel-Plana}, Appendix \ref{sec-anotherExpressionSd}: An example of
the correlation functions with
discrete spectrum is calculated and discussed, with the use of the Abel-Plana formula.
The temperature $T_{rot}$ associated with rotation is introduced.
Section \ref{sec:EM_EnergyDensity}. Expression for energy density of the random classical zero-point electromagnetic field measured by a rotating detector is constructed. It is explicitly
shown to display thermal features, following spectrum discreteness
observed by the detector.
Section \ref{sec-massless} is dedicated to detector
rotation in massless zero-point scalar field radiation.
Subsection \ref{sec:scalarCF}: A correlation function
of the massless zero-point scalar field is calculated with the use of tetrad formalism.
Subsection \ref{sec:PeriodictyScalaeCF}: The correlation function of the massless zero-point scalar field for a discrete spectrum  following its periodicity is defined.
Its spectrum contains the Planck's factor.  Subsection \ref{sec:scalarEnergyDensity}:
Energy density of the massless scalar field measured by a rotating detector, and their thermal properties connected with the detector rotation and periodicity are obtained and discussed.
Section \ref{sec:Discusssion}: Conclusion and Perspectives.
\section{ Electromagnetic Field at a Rotating
Detector Moving Through a Random Classical Zero - Point Radiation. }\label{sec-CorFunForEelctromagneticFieldAtRot}
\subsection{  Local Measurements, Tetrads, and Correlation Functions.  }
\label{sec:TetradsMeasuments}
\indent Let the detector  be a particle moving through an electromagnetic field in Minkowsky space-time and the detector measures it
on the world line in a locally inertial reference frame. We assume that the field is  classical
and in a vacuum state. Mathematical definition of the vacuum field state is given in the next subsection. \\
\indent The  quantities associated with
such local measurements  can be
 described in 4-orthogonal tetrad (OT) formalism \cite{synge}, \cite{Landau}.
\indent  Any vector or  tensor may be resolved along 4 tetrad vectors $\mu^i_{(a)}$, a =1, 2, 3, and 4.  ( The tetrad vectors are described in  Appendix \ref{sec:Fermi-Walker} ). For example, a 4-vector velocity of
a detector and the tensor of electromagnetic field are respectively
\begin{eqnarray}
U_i =U_{(a)}(\mu) \; \mu^{(a)}_i, \;\;\;  F_{ik}=\mu^{(a)}_i \; \mu^{(b)}_k \; F_{(ab)}(\mu).
\end{eqnarray}
The components
\begin{eqnarray}
 U_{(a)}(\mu)=\mu^{i}_{(a)} U_i
\end{eqnarray}
and
\begin{eqnarray}
\label{eq:FrameTensor} F_{(ab)}(\mu)=\mu^{i}_{(a)} \; \mu^k_{(b)}
\; F_{i k}.
\end{eqnarray}
 are invariants in the tensorial sense (i.e with respect to coordinate transformation) and defined in a local reference frame with locally lorentz-invariant metrics tensor
 $\eta_{ab}=\eta^{ab}=diag(1,1,1,-1)$ (Appendix \ref{sec:Fermi-Walker} ) .
Therefore they   describe local observable quantities. \\
\indent In this work  OTs are
defined as Frenet-Serret orthogonal tetrads associated with each point of the world line of
the rotating detector with  4-vector velocity
\begin{eqnarray}
U^i=c\;(-\beta \gamma \sin \alpha,\beta \gamma \cos \alpha, 0,
\gamma),
\end{eqnarray}
where  $\beta=
 v/c=\Omega a /c, \gamma=(1-\beta^2)^{-1/2}, \alpha= \Omega
 \gamma \tau$, and $\Omega, \;a$ are  angular velocity and
circumference radius  of the rotating detector respectively.
4-vectors of Frenet-Serret OT, solutions of the equations (\ref{eq: Frenet-Serret_eqns}), have the form:
\begin{eqnarray}
\label{eq:FStetrad}
 \mu^i_{(4)}= \frac{U^i}{c}, \nonumber \\
\mu^i_{(1)}=(\cos \alpha, \sin \alpha, 0,0), \nonumber \\
\mu^i_{(2)}=(-\gamma \sin \alpha, \gamma \cos \alpha,0, \beta \gamma).
 \nonumber \\
 \mu^i_{(3)}=(0,0,1,0).
\end{eqnarray}
In  local reference frames, defined by these tetrads, the
detector is at rest:
\begin{eqnarray}
U_{(a)}=\mu^i_{(a)}U_i= \mu^i_{(a)} U^k g_{ik}= (0,0,0,-c).
\end{eqnarray}
The 3-vector acceleration of the detector in it is constant in both magnitude and direction:
\begin{eqnarray}
\dot{U}_{(a)}= \mu^i_{(a)}\dot{U}_i=\mu^i_{(a)}\dot{U}^k g_{ik}
=(-a \Omega^2 \gamma^2,0,0,0), \;\;\; g_{ik}=diag(1,1,1,-1),
\end{eqnarray}
as it would be in the case of a uniformly accelerated detector. It is why we preferred to use Frenet-Serret tetrads, and  not Fermi-Walker ones.
Fermi-Walker tetrads  do not have this feature ( see Appendix
\ref{sec:Fermi-Walker}). \\
\indent Following  formulas (\ref{eq:FrameTensor}) and
( \ref{eq:FStetrad}),  the electric
 $ E_{(k)}(\mu | \tau)$ and magnetic  $H_{(k)}(\mu | \tau)$
 fields, which denote local observable quantities,  in the Frenet-Serret reference frame  $\mu_{\tau}$ at the proper time
$\tau$ of the rotating detector
can be given in
terms of electric $E_k$ and magnetic $H_k$ fields in the inertial laboratory
coordinate system:
\begin{eqnarray}
\label{eq:FieldAtTetrad} E_{(1)}(\mu | \tau)=F_{(41)}(\mu | \tau)=
E_1 \gamma \cos \alpha + E_2\gamma \sin \alpha
-H_3 \beta \gamma, \nonumber \\
E_{(2)}(\mu | \tau)=F_{(42)}=E_1(-\sin \alpha) +
E_2\cos \alpha, \nonumber\\
E_{(3)}(\mu | \tau)=F_{(43)}= E_3 \;\;\gamma + H_1\beta \gamma
\cos\alpha +
H_2 \beta \gamma \sin \alpha,  \nonumber \\
H_{(1)}(\mu | \tau)=F_{(23)}= H_1\gamma \cos \alpha + H_2\gamma
\sin \alpha
+ E_3\beta \gamma,  \nonumber \\
H_{(2)}(\mu | \tau)=F_{(31)}= H_1(-\sin \alpha) +
H_2 \cos \alpha,\nonumber  \\
H_{(3)}(\mu | \tau)=F_{(12)}= H_3\gamma + E_1(-\beta \gamma \cos
\alpha) + E_2(-\beta \gamma \sin \alpha),
 \end{eqnarray}\\
where $\alpha=\Omega \gamma \tau$. \\
\indent A mathematical subject of  this work is bilinear combinations of the local
fields, which are taken  in two tetrads, averaged over the field in a vacuum state defined in the
laboratory coordinate system. Formulas (\ref{eq:FieldAtTetrad}) can be used to calculate the following
 two-field correlation functions (CF) of the electromagnetic field at the rotating detector:
 \begin{eqnarray}
 \label{eq: CF}
 I_{(ab)}^E \equiv \langle E_{(a)}(\mu_1|\tau_1)E_{(b)}(\mu_2|\tau_2)\rangle, \;\;
 I_{(ab)}^{EH} \equiv \langle E_{(a)}(\mu_1|\tau_1)H_{(b)}(\mu_2|\tau_2)\rangle, \;\;
 I_{(ab)}^H \equiv\langle H_{(a)}(\mu_1|\tau_1)H_{(b)}(\mu_2|\tau_2)\rangle,
 \end{eqnarray}
 where $a,b=1,2,3$. In these expressions $\mu_1$ and $\mu_2$ are two reference frames
(tetrads) on the circle of the rotating detector at the proper
times $\tau_1$ and $\tau_2$ respectively. For example,
\begin{eqnarray}
\label{eq: CF1}
I_{(11)}^E= <E_1(\tau_1)E_1(\tau_2)> \gamma^2 \cos \alpha_1 \cos
\alpha_2 + <E_1(\tau_1)E_2(\tau_2)> \gamma^2 \cos \alpha_1 \sin
\alpha_2 +  \nonumber \\
<E_2(\tau_1) E_1(\tau_2)> \gamma^2 \sin \alpha_1 \cos \alpha_2 +
<E_1(\tau_1)H_3(\tau_2)> (-1)\beta \gamma^2 \cos \alpha_1 +
\nonumber \\
<H_3(\tau_1)E_1(\tau_2)>(-1)\beta \gamma^2 \cos \alpha_2 +
<E_2(\tau_1)E_2(\tau_2)> \gamma^2 \sin \alpha_1 \sin \alpha_2 +
\nonumber \\
<E_2(\tau_1)H_3(\tau_2)>(-1) \beta \gamma^2 \sin \alpha_1 +
<H_3(\tau_1)E_2(\tau_2)>(-1)\beta \gamma^2 \sin \alpha_2 +
\nonumber \\
(\beta \gamma)^2 <H_3(\tau_1)H_3(\tau_2)>.
\end{eqnarray}
The expressions for some other CFs are given in Appendix \ref{sec-generalExpressions}. They follow  from (\ref{eq:FieldAtTetrad}).
When $\tau_1 \rightarrow \tau_2$ these expressions can be used to calculate
expectation values for energy density.
Here $\langle \rangle$ means averaging over a vacuum state of the electromagnetic field in the
laboratory coordinate system. In the next section, we will consider averaging for a situation
when a vacuum state of the electromagnetic field is a random classical zero point radiation.
\subsection{Correlation Function Calculation Scheme: Example for $I^E_{(11)}$.} \label{sec-IntegralExpressions}
\indent In the classical case,  the electric and
magnetic field components $E_k$ and $H_k$ in (\ref{eq:FieldAtTetrad}) and (\ref{eq: CF1})
represent  the random zero-point radiation in the
laboratory coordinate system
\cite{boyer1980}(47), (48) at a time-space position $(t, \vec{r})$ of the rotating detector :
\begin{eqnarray}
\label{eq:FieldOnCircle} \vec{E}(\tau)= \sum^2_{\lambda=1} \int
d^3k
\hat{\epsilon}(\vec{k},\lambda)h_0(\omega)\cos[\vec{k}\vec{r}(\tau)-
\omega
\gamma \tau -\theta(\vec{k},\lambda)], \nonumber \\
\vec{H}(\tau)=\sum^2_{\lambda=1} \int d^3k
[\hat{k},\hat{\epsilon}^(\vec{k},\lambda)]h_0(\omega)\cos[\vec{k}
\vec{r}(\tau)-\omega \gamma \tau-\theta(\vec{k},\lambda)],
\label{eq:ff}
\end{eqnarray}
where, in distinction from \cite{boyer1980}, the laboratory coordinates $\vec{r}(t)$ and time t are taken in terms of
proper time $\tau$ of the rotating observer:
\begin{eqnarray}
\label{eq:radius_vector}
\vec{r}(\tau)=
 (a \cos \Omega \gamma \tau, a \sin \Omega \gamma \tau, 0), \;\;\; t =\gamma \tau,
\end{eqnarray}
the  $\theta(\vec{k},\lambda)$ describe random phases distributed
uniformly on the interval $(0,2\pi)$ and independently for each
wave vector $\vec{k}$ and polarization $\lambda$ of of a plane
wave, and
\begin{eqnarray}
 \pi^2 h_0^2(\omega)=(1/2)\hbar \omega .
 \end{eqnarray}
\indent Averaging $\langle \rangle$ in (\ref{eq: CF1}) means averaging over random phases $\theta(k, \lambda)$.
To illustrate a technique of CF calculation, we will compute the CF $I_{(11)}^E$ as
an example. This technique is very similar to one in \cite{boyer1980} developed
for a uniformly accelerated case, as apposed to rotation,  though the tetrad formalism is not
used there.\\
\indent The $<>$ expressions in ( $\ref{eq: CF1}$ ),  contain double integrals and double sums
$\int d \vec{k} \int d \vec{k}^{\prime} \sum_{\lambda} \sum_{\lambda^{\prime}}$. Using the known
$\theta$ - function properties \cite{boyer1980}
\begin{eqnarray} \label{eq:thetas}
<\cos \theta(\vec{k}, \lambda)\cos \theta(\vec{k}^\prime,
\lambda^\prime)>=<\sin \theta(\vec{k}, \lambda)\sin
\theta(\vec{k}^\prime, \lambda^\prime)>= \frac{1}{2}
\delta_{\lambda \;\lambda^\prime}
\delta^3(\vec{k}-\vec{k}^\prime), \nonumber \\
<\cos \theta(\vec{k}, \lambda)\sin
\theta(\vec{k}^\prime, \lambda^\prime)>=0
\end{eqnarray}
and the sum over polarization
\begin{eqnarray}
\sum^2_{\lambda=1}\epsilon_i(\vec{k},
\lambda)\epsilon_i(\vec{k}^\prime,
\lambda^\prime)=\delta_{ij}-k_i\;k_j/k^2 \equiv \delta_{ij}-\hat{k}_i
\hat{k}_j,
\end{eqnarray}
 they can be reduced to an integral-sum of the the type
$\int d \vec{k} \sum_{\lambda}$. Then using variable change in the integrands, from $\vec{k}$
to $\vec{k}^{\prime}$,
\begin{eqnarray}
\label{eq:VariableChange} \hat{k}_x \cos \alpha + \hat{k}_y \sin
\alpha =\hat{k^\prime}_x, \;\; -\hat{k}_x \sin \alpha + \hat{k}_y
\cos \alpha =\hat{k^\prime}_y,
\end{eqnarray}
with
\begin{eqnarray}
\alpha=\frac{\alpha_1 +
\alpha_2}{2}=\frac{\Omega \gamma (\tau_2+\tau_1)}{2}, \;\;
\hat{k}_i=k_i/k, \;\; i=x, y, z,
\end{eqnarray}
we come to the following expressions for the $<>$ terms in (\ref{eq: CF1}):
\begin{eqnarray}
<E_1(\tau_1)E_1(\tau_2)>=\int d^3 k \;R + (-\cos^2\alpha) \int
d^3k \; \hat{k}^2_x\; R + (-\sin^2\alpha) \int d^3k \;
\hat{k}^2_y\; R, \nonumber \\
<E_1(\tau_1)E_2(\tau_2)>=<E_2(\tau_1)E_1(\tau_2)>= -\frac{\sin 2
\alpha}{2}\int d^3\;k \; \hat{k}^2_x \;R + \frac{\sin 2
\alpha}{2}\int d^3\;k \; \hat{k}^2_y \;R , \nonumber \\
<E_1(\tau_1)H_3(\tau_2)>=<E_1(\tau_2)H_3(\tau_1)>= -\cos\alpha
\;\int d^3\;k \;\hat{k}_y \; R, \nonumber \\
<E_2(\tau_1)E_2(\tau_2)>=\int d^3\;k\;R + (-\sin^2 \alpha)\int d^3
k \; \hat{k}^2_x\; R + (-\cos^2 \alpha)\int d^3 k \; \hat{k}^2_y\;
R, \nonumber \\
<E_2(\tau_1)H_3(\tau_2)>= <E_2(\tau_2)H_3(\tau_1)>= (-\sin \alpha)
\int d^3 k\; \hat{k}_y \; R, \nonumber \\
<H_3(\tau_1)H_3(\tau_2)>= \int d^3 k \; \hat{k}^2_x \;R+ \int d^3
k \; \hat{k}^2_y \;R.
\end{eqnarray}
In these expressions, the prime symbol of the ``dummy" variable
$k^{\prime}$ is omitted for simplicity, and we use the following notations:
\begin{eqnarray}
\label{eq-expressionF}
R=h^2_0(\omega) \;\frac{1}{2}\cos k F , \;\;\; F=c \gamma
(\tau_2-\tau_1)[1-\hat{k}_y \frac{v}{c}\frac{\sin
\delta/2}{\delta/2}],\;\;
 \delta=\alpha_2-\alpha_1=\Omega \gamma
 (\tau_2-\tau_1).
\end{eqnarray}
\indent After some  simplifications we come to the following expression for
$I_{(11)}^E$:
\begin{eqnarray}
\label{eq:CF11}
I_{(11)}^E=\langle E_{(1)}(\mu_1|\tau_1)E_{(1)}(\mu_2|\tau_2)
\rangle=\gamma^2 \cos \delta \int d^3k \;h^2_0(\omega) \frac{1}{2}
\cos kF + 2 \beta \gamma^2 \cos \frac{\delta}{2} \int d^3 k
\;\hat{k}_y \; h^2_0(\omega)\frac{1}{2} \cos kF + \nonumber
\\
\gamma^2 [\beta^2 - \cos^2 \frac{\delta}{2}]\int d^3 k\;
\hat{k}^2_x \; h^2_0(\omega)\; \cos kF+ \gamma^2[\beta^2 +
\sin^2\frac{\delta}{2}] \int d^3 k\; \hat{k}^2_y
h^2_0(\omega)\frac{1}{2} \cos kF. \nonumber \\
\end{eqnarray}
\indent This function clearly depends only on the proper time
interval $\tau_2-\tau_1$ and is not dependent on  $(\tau_1+\tau_2)/2$ that is
\begin{eqnarray}
I_{(11)}^E=I_{(11)}^E(\tau_2 - \tau_1).
\end{eqnarray}
 General expressions for other
CFs can be found in  Appendix \ref{sec-generalExpressions}. They
have the same properties and also depend only on the proper time
interval $\tau_2-\tau_1$.
\subsection{
\textbf{The Correlation Function $I_{(11)}^E$ } in Terms of Elementary
Functions.} \label{sec-Final}
\indent The CF  $I_{(11)}^E \equiv \langle
E_{(1)}(\mu_1|\tau_1)E_{(1)}(\mu_2|\tau_2) \rangle$ defined and discussed above can be
represented in terms of elementary functions.
After integration of (\ref{eq:CF11}) in spherical coordinates, over k and
then over $\phi$, we come to the expression:
\begin{eqnarray}
\label{eq-CF11Final}
I_{(11)}^E = \frac{3\hbar
c}{2 \pi^2 [c(t_2-t_1)]^4} \gamma^2 \{ + [2 \pi \cos \delta]
\int_{0}^{\pi}d\theta\frac{\sin\theta}
{(1-k^2\sin^2\theta)^{7/2}} \nonumber \\
+[3\pi k^2 \cos\delta -2 \pi \cos^2(\delta/2)+ 2\pi \beta^2 - 8
\pi \beta k \cos(\delta/2)+\pi]
\int_{0}^{\pi}d\theta\frac{\sin^3\theta}
{(1-k^2\sin^2\theta)^{7/2}} \nonumber \\
+[-3 \pi k^2 \cos^2(\delta/2) + 3 \pi \beta^2 k^2 -2 \pi \beta k^3
\cos(\delta/2) + 4 \pi k^2]
\int_{0}^{\pi}d\theta\frac{\sin^5\theta}
{(1-k^2\sin^2\theta)^{7/2}} \}
\end{eqnarray}
(see  Appendix \ref{sec-IntegralCalculationFinal} for details).
The integrals over $\theta$ in this expression are :
\begin{eqnarray}
\label{eq-CF11-0Final} \int_{0}^{\pi}d\theta\frac{\sin\theta}
{(1-k^2\sin^2\theta)^{7/2}}= \frac{2}{5(1-k^2)} +
\frac{8}{15(1-k^2)^2} + \frac{16}{15(1-k^2)^3} , \label{eq:sin1} \\
 \int_{0}^{\pi}d\theta\frac{\sin^3\theta}
{(1-k^2\sin^2\theta)^{7/2}}= \frac{4}{15(1-k^2)^2}
+\frac{16}{15(1-k^2)^3}, \label{eq:sin3}\\
\int_{0}^{\pi}d\theta\frac{\sin^5\theta}
{(1-k^2\sin^2\theta)^{7/2}}=\frac{16}{15(1-k^2)^3}
 \label{eq:sin5}.
\end{eqnarray}
(See formulas \cite{prudnikov}, 1.5.23, 1.2.43. )
In these expressions, k is not a module of a wave vector $\vec{k}$, but a constant for the CFs:
\begin{eqnarray}
 k=-\frac{v}{c}\frac{\sin \delta/2}{\delta/2}, \; \; \delta=\Omega \gamma (\tau_2 - \tau_1). \nonumber
\end{eqnarray}
\indent Other CFs  can also
be expressed in  terms of elementary functions.\\
\indent In this form the CFs do not display thermal features. In
the next section we will investigate under what conditions they can display thermal
properties. We will show that periodic CFs have thermal features.
\subsection{Periodicity of  Correlation Functions: Example for $ I_{(11)}^E $. }
\label{sec:PeriodicityOfCF}
\indent We assume that CFs at a rotating detector should be
periodic because CF measurements is one of the tools the detector
can use to justify the periodicity of its motion.  Mathematically
it means that
\begin{eqnarray}
 I_{(11)}^E(t_2 - t_1) = I_{(11)}^E(\;\;(t_2 - t_1)+ \frac{2\pi}
 {\Omega}n \;\;)
\end{eqnarray}
or
\begin{eqnarray}
 I_{(11)}^E(\tau_2 - \tau_1) = I_{(11)}^E(\;\;(\tau_2 - \tau_1)+ \frac{2\pi}
 {\Omega \gamma}n \;\;)
\end{eqnarray}
  Here
$\Omega=\frac{2\pi}{T}$ is an angular velocity of the rotating
detector and $n=  0, 1, 2, 3,...$ . Breaking down $\cos kF$ in (\ref{eq:CF11}) into odd and
even powers of  $k_y$ and taking into consideration that the odd part of the integrand gives zero after  integration over $k_y$ it is easy to show that the CF is periodic  if in its integrand
\begin{eqnarray}
\omega=c k =\Omega n.
\end{eqnarray}
 It means that the rotating detector observes not the entire random
 electromagnetic radiation spectrum but only a discrete part of it.
We could also expect the same result based  on the following consideration.
Even though no assumptions about a structure of the rotating detector have been made
so far, it should have some common features connected with the type of its motion.
First of all it should have a
charge simply  because a neutral, not charged, detector cannot be used to observe
electromagnetic field and can not be kept on a circular orbit. Then the charge of the rotating detector behaves as an
oscillator with a frequency $\Omega$ and resonance frequencies $ n \Omega$.
Of course  this discrete spectrum  is the same as the radiation spectrum of a rotating electrical charge \cite{ivanenko}(39.29). \\
\indent The  expression (\ref{eq-CF11Final})  for $I_{(11)}^E$
cannot be used to analyze the periodicity consequences because
the integration over entire continuous spectrum of $\omega$ has
already been done in it. It is why we have used the expression
(\ref{eq:CF11}), before the integration over $\omega$. \\
 \indent Let us now consider the  correlation
function $I_{(11)}^E$, periodic over $\tau$, with the discrete spectrum. There are
two ways to do this. The first one is simpler, just to modify the formula
(\ref{eq:CF11}) for $I_{(11)}^E$ for the discrete spectrum. It will
be described below in the next subsection. The  second  one is
identical with the approach we have used above for the continuous
spectrum but with the modified equations (\ref{eq:ff}) and
relationships (\ref{eq:thetas}) for the discrete spectrum. It is
described in  Appendix \ref{sec-ModifiedExpressionRandom}.
\subsection{ Correlation Functions \textbf{With
the Discrete Spectrum}: Example for  $ I_{(11)}^E $.  }
\label{sec:CFDiscrete} The integrals in (\ref{eq:CF11}) can be
represented as
\begin{eqnarray}
\label{eq:CF_Integral} \int d^3 k [ \;\; ] \frac{1}{2} h^2_0
(\omega) \cos kF =
 \frac{c \hbar k^4_0}{4 \pi^2}\int d O [\;\;] S,
\end{eqnarray}
where
\begin{eqnarray}
S=\int d \kappa \;\kappa^3 \cos \kappa F_d , \;\;\; d O =d\theta
d\phi \sin \theta, \;\;\; \kappa=\frac{k}{k_0}, \;\; k_0 = \Omega /c,
\end{eqnarray}
and
\begin{eqnarray}
 F_d=k_0 F= \delta [1 - \hat{k}_y \frac{v}{c}\frac{\sin \delta /2}
 {\delta/2}],
\end{eqnarray}
The expressions in $[\;\;\;]$ are 1, $\hat{k}_y =\frac{k_y}{k}$, $\hat{k}_x^2
= (\frac{k_x}{k})^2$, and $\hat{k}_y^2 = (\frac{k_y}{k})^2$  do not depend on
$\kappa$. \\
\indent For the discrete spectrum case the integration in
(\ref{eq:CF_Integral}) over $\kappa$ should be changed to
summation over n. So the the only term to be changed is S $\rightarrow S_d $. It
becomes
\begin{eqnarray}
\label{eq:CF11d2} S_d=\sum_{0}^{\infty} n^3 \cos n F_d.
\end{eqnarray}
Then the periodical CF, corresponding to (\ref{eq:CF11}),  with
the discrete spectrum  can be defined in the form
\begin{eqnarray}
\label{eq:CF11d1}
I_{(11)d}^E \equiv \langle E_{(1)}(\mu_1|\tau_1)E_{(1)}(\mu_2|\tau_2) \rangle_d=  \frac{c
\hbar k^4_0}{4 \pi^2} \;\; \{ \;\;\gamma^2 \cos \delta \int d O \;
S_d+ 2 \beta \gamma^2 \cos \frac{\delta}{2} \int dO \;\hat{k}_y
\;S_d
 + \nonumber \\
\gamma^2 [\beta^2 - \cos^2 \frac{\delta}{2}]\int d O\; \hat{k}^2_x
\; S_d+ \gamma^2[\beta^2 + \sin^2\frac{\delta}{2}] \int d O\;
\hat{k}^2_y S_d \;\;\},
\end{eqnarray}
where integration is held on angular variables only, and $S_d$ is a series sum which is analyzed
in the next section using  the Abel-Plana
formula.
\subsection{ The Abel-Plana Formula and Thermal Properties of  Correlation Functions With the Discrete Spectrum: Example for $I^E_{(11)d}$. }
\label{sec:Abel-Plana}
\indent Using Abel-Plana summation formula
\cite{Bateman1953},\cite{MT1988}, \cite{Evgrafov1968}
\begin{eqnarray}
\label{eq:AbelPlana}
 \sum_{n=0}^\infty \, f(n)= \int_0^\infty
f(x)\,dx + \frac{f(0)}{2} +i \,\int_0^\infty \,dt \,
\frac{f(it)-f(-it)}{e^{2 \pi t}-1},
\end{eqnarray}
with
\begin{eqnarray}
f(n)= n^3 \cos n F_d
\end{eqnarray}
 we come to the following expression
for $S_d$ (\ref{eq:CF11d2}):
\begin{eqnarray}
\label{eq:sum}
 \Omega^4 S_d
 =  \int _0^{\infty} d \, \omega \omega^3 \cos( \omega \tilde{F})+
 \int_0^{\infty} d\omega \frac{2 \omega^3 \cosh(\omega \tilde{F})}{e^
 {2\pi \omega/\Omega}-1}, & \tilde{F}=\frac{F_d}{\Omega},
\end{eqnarray}
and the CF (\ref{eq:CF11d1}) becomes
\begin{eqnarray}
 \label{eq:CF11d3}
 I_{(11)d}^E =\langle E_{(1)}(\mu_1|\tau_1)E_{(1)}(\mu_2|\tau_2) \rangle_d=
  \int d O \;\;  K(\theta, \phi, \delta) \; \times \nonumber \\  \frac{2}{3}\;\frac{ \hbar}{ \pi c^3} \; \{ \int _0^{\infty} d \, \omega \omega^3 \cos( \omega \tilde{F})+
 \int_0^{\infty} d\omega \frac{2 \omega^3 \cosh(\omega \tilde{F})}{e^
 {2\pi \omega/\Omega}-1} \},
 \end{eqnarray}
 where
\begin{eqnarray}
K(\theta, \phi, \delta) = \frac{3}{8 \pi}
\;\;\{ \;\;
\gamma^2 \cos \delta \;  +
2 \beta \gamma^2 \cos \frac{\delta}{2} \;\hat{k}_y
\;  +
\gamma^2 [\beta^2 - \cos^2 \frac{\delta}{2}]\; \hat{k}^2_x
\; + \gamma^2[\beta^2 +
 \sin^2\frac{\delta}{2}]\; \hat{k}^2_y.
\end{eqnarray}
\indent Expressions for $S_d$ after integration over $\omega$ are given in (\ref{eq:I_one}), Appendix
\ref{sec-anotherExpressionSd}, and further discussion could have been made in terms
of obtained elementary functions. But it is simpler to consider the structure of the integrand in the expression for $S_d$ explicitly. \\
\indent The CF  $I^E_{(11)d}$
resembles the CF of the thermal radiation, with Planck's spectrum and zero-point radiation included,
observed by a detector at rest in an
inertial frame \cite{boyer1980}(73) :
\begin{eqnarray}
\label{eq:Tspectral function}
\langle E_{Ti}(0,s - t/2) E_{Ti}(0, s + t/2) \rangle=
\frac{2}{3}\frac{\hbar}{\pi c^3} \{\int_0^{\infty}d\omega\omega^3 \cos \omega t +
\int_0^{\infty}d \omega \frac{2 \omega ^3 \cos \omega t}{ e^{\frac{\hbar \omega}{k T} }-1 }  \}
\end{eqnarray}
which corresponds to the spectral function
\begin{eqnarray}
\pi^2 \hbar^2_{T}(\omega) =\frac{1}{2}\hbar \omega \coth \frac{\hbar \omega}{2 k T}=
\hbar \omega (\frac{1}{2} + \frac{1}{e^{\hbar \omega /k T}-1}).
\end{eqnarray}
\indent Indeed, from (\ref{eq:CF11d3}) and (\ref{eq:Tspectral function}) we can see that
the integrands in both expressions  have the Planck factor $1/(\exp^{\frac{\hbar
\omega}{k_B}T}-
1 )$ if, in  (\ref{eq:CF11d3}), we define  a new constant, a rotation temperature,
$T_{rot}$ as
\begin{equation}
\label{eq:T}
 T_{rot}=\frac{\hbar \Omega}{2 \pi k_B}.
\end{equation}
The Planck  factor is an indication that some thermal effects accompany
the detector rotation in the random classical zero-point
electromagnetic radiation  though there is also a significant distinction
between them. In (\ref{eq:CF11d3}),
\emph{$\tilde{F}=t(1-\hat{k}_y  \frac{v}{c}\frac{\sin (\Omega t/2) }{\Omega t/2})$}
and \emph{cosh} are used instead of \emph{t} and \emph{cos}
respectively in (\ref{eq:Tspectral function}).  The coefficient
$\tilde{F}$ depends on both $\theta$ and $\phi$ because $\hat{k}_y=\sin
\theta \sin\phi$. Besides the expression (\ref{eq:CF11d3}), compared to (\ref{eq:Tspectral function}), contains coefficient $K(\theta, \phi, \delta)$ and integration over $\theta$
and $\phi$.\\
\indent So the CF $I^E_{(11)d}$ at a rotating detector explores some thermal properties but
 does not coincide with the CF ($\ref{eq:Tspectral function}$) at an inertial observer put in the radiation with  Planck's radiation. Partly it occurs  because  radiation, isotropic in the laboratory system, looks anisotropic for a rotating detector.
Is there any situation when  operands in ($\ref{eq:CF11d3}$) and ($\ref{eq:Tspectral function}$)  are identical ? It is easy to see that in the limit   $ t \rightarrow 0 $  and therefore $\tilde{F}\rightarrow 0$ , when two observation
 points, $\tau_1$ and $\tau_2$ (or $t_1$ and $t_2$ in the laboratory system) coincide,  both expressions are identical. This observation brings up the
idea that the energy density ( one-observation-point quantity and consisting of diagonal elements of the CF ) of
the random classical electromagnetic radiation measured by a
detector, rotating through a zero point radiation, has the Planck
spectrum at the temperature $T_{rot}$ (\ref{eq:T}). This issue
will be discussed in the next section.
\section{  The Energy Density of Random Classical
Electromagnetic Radiation Observed by a Rotating detector: Periodicity and
Planck's Spectrum.}
\label{sec:EM_EnergyDensity}
In any reference frame $\mu_{\tau}$, with Minkowsky metrics $\eta_{(ab)}$, local lorentz coordinates can be introduced \cite{moller72}, section 9.6. The local
reference frame, defined this way, is an inertial system, and all laws of Special Relativity should be true in this locally inertial reference frame. Then the energy density measured by the rotating observer at $\mu_{\tau}$ will be of the form:
\begin{equation}
\label{eq:energy density}
w =\frac{1}{8 \pi} \sum_{a=1}^3\;( \;\;\langle
E_{(a)}^2(\mu|\tau)\rangle + \langle H_{(a)}^2(\mu|\tau)\rangle
\;\; )
\end{equation}
or, in terms of electric and magnetic fields measured in
the laboratory coordinate system (\ref{eq:FieldAtTetrad}),
\begin{equation}\label{eq:energy density}
w =\frac{1}{4\pi}\{\;\;[\;\langle E_1^2 \rangle + \langle
E_3^2 \rangle \; ]\gamma^2(1+\beta^2)+\langle E_2^2 \rangle \;\;\} + \nonumber \\
\frac{1}{8 \pi} 4 \; \gamma^2 \beta \;
(\;\langle E_1 H_3 \rangle - \langle E_3 H_1 \rangle \;),
\end{equation}
where as we will show below $\langle E_i^2 \rangle =\langle H_i^2
\rangle $, $i=1,2,3$, and $w$ does not depend on the choice of a tetrad $\mu$. \\
\indent We have already seen that the correlation functions with a periodicity have a discrete spectrum.
Effectively, in calculations, it means that integral expressions for zero-point random radiation fields $E_i$ and $H_i$ in the laboratory coordinates should be modified and presented as
series over frequencies.
Explicit expressions for the fields $E_i$ and $H_i$ with discrete spectrum are given in
Appendix \ref{sec-ModifiedExpressionRandom} and could be used in ( $\ref{eq:energy density}$) to take into consideration periodicity.
With the help of these formulas  and using the technique for discrete spectrum
described above  we come to the following
expressions
\begin{eqnarray}
\langle E_1 H_3 \rangle - \langle E_3 H_1 \rangle =0,
\end{eqnarray}
and
\begin{eqnarray}
\label{eq:FiledSquared} \langle E_i^2 \rangle = \langle H_i^2
\rangle = \frac{k_0^4 \hbar c}{2 \pi^2} \int dO (1-\hat {k}^2_i
)\sum^\infty_{n=0}n^3, \;\;\; k_0 = \Omega /c
\end{eqnarray}
for $i=1,2,3$. Finally, after integration over $\theta$ and $\phi $, we have
\begin{equation}
\label{eq: energy density 1} w = \frac{ (4\gamma^2-1)}{3
}\;\frac{\hbar}{c^3 \pi^2}\; \Omega^4\sum_{n=0}^\infty n^3.
\end{equation}
Using the Abel-Plana formula
(\ref{eq:sum}) with $F_d=0$ this expression can be given in the form:
\begin{equation}
\label{eq:energy density 2} w = \frac{2 \;(4\gamma^2-1)}{3 } \;\; (w_{ZP} + w_T ),
\end{equation}
where
\begin{eqnarray}
\label{eq: energy density 3}
w_{ZP} =\frac{\hbar }{c^3
\pi^2} \; \; \int _0^{\infty} d \, \omega \frac{1}{2}\omega^3, \;\;\;
w_T = \frac{\hbar }{c^3 \pi^2} \; \; \int_0^{\infty} d\omega \frac{ \omega^3 }
 {e^{\hbar\omega/k_B T_{rot}} -1} = 4 \; \frac{\pi^2 k_B^4}{60 (c \hbar)^3}\; T_{rot}^4=
\frac{4 \sigma}{c} T_{rot}^4,
\end{eqnarray}
$k_B$ is the Boltzman constant, and $\sigma$ is the Stefan-Boltzman constant.\\
\indent Thus, due to the periodicity of
the motion, the detector rotating in the zero-point  radiation
under the temperature $T=0$  observes not only  original
zero-point radiation, $w_{ZP}$, but also the radiation, $w_T$, with Planck's spectrum if parameter $T_{rot}$ is interpreted as the
temperature associated with the detector rotation. Expression $w_T$ is exactly the energy density of the black radiation at the temperature $T_{rot}$ [\cite{landau_lifshits}, (60,14)]. The factor
$\frac{2}{3}(4\gamma^2-1)$ comes from integration in (\ref{eq:FiledSquared}) over
angles due to anisotropy of the electromagnetic field measured by the rotating observer.\\
\indent All this consideration is true when
$\Omega r < c $. The first term  of (\ref{eq:energy density 2}), corresponding to ZP
radiation,  is divergent for any r and $\Omega$. The second one, describing  the thermal properties, is convergent, though it is growing to infinity if $r \rightarrow c / \Omega $  for a fixed $\Omega $ or $\Omega \rightarrow c/r$ for a fixed $ r $.
\section{Random Classical Massless Zero-Point Scalar Field at a
Rotating Detector. } \label{sec-massless}
\subsection{ Correlation Function.}
\label{sec:scalarCF}
\indent The scalar field $\psi_s(\mu_{\tau}|\tau)$ in a tetrad $\mu_\tau$
 has the same form as in the laboratory
coordinate system, $\psi_s(\tau)$ , taken in the location of the
tetrad, because it is a scalar. Then the correlation function
measured by an observer rotating through a random classical massless
zero-point scalar field radiation has the form  \cite{boyer1980}:
\begin{eqnarray}
\langle \psi_s(\mu_1|\tau_1) \psi_s(\mu_2 |\tau_2)\rangle =
\langle \psi_s(\tau_1) \psi_s(\tau_2) \rangle ,
\label{eq:CF_Scal_Field_Continuous}
\end{eqnarray}
where
\begin{eqnarray}
\label{eq:scalarField} \psi_s(\tau_i)= \int d^3k_i f(\omega_i)
\cos \{ \vec{k}_i \vec{r}(\tau_i) -\omega_i \gamma \tau_i -
\theta(k_i)\},
\end{eqnarray}
and (instead of $\hbar_0(\omega)$  in (\ref{eq:ff}) )
\begin{eqnarray}
\label{eq:scalarTheta}
   f^2(\omega_i)=\frac{\hbar c^2}{2 \pi^2 \omega_i}, \;\;  \omega_i =c k_{0i},
  \;\;\;\; i=1,2 .
\end{eqnarray}
The $\theta$-functions,  $\vec{r}(\tau_i)$, and $t(\tau_i)$ are defined in (\ref{eq:radius_vector}) and (\ref{eq:thetas}).
Using these expressions and variable change (\ref{eq:VariableChange})
in the double-integral (\ref{eq:CF_Scal_Field_Continuous}) we get  the expression :
\begin{eqnarray}
\label{eq:ScalarCF}
 \langle \psi_s(\mu_1 |\tau_1) \psi_s(\mu_2 |\tau_2)\rangle=
 \int
d^3 k f^2(\omega)\frac{1}{2} \cos kF,
\end{eqnarray}
where F is defined in (\ref{eq-expressionF}).
 Having integrated over k, $\phi$ , and $\theta$ we come to the  expression
 for the CF of the random classical
 massless
 scalar field at the rotating  detector moving through a zero point
massless scalar radiation ( see details in Appendix \ref{sec:CF_Scalar_Field} ):
\begin{eqnarray}
\label{eq:WightmanFunction}
\langle \psi_s(\mu_1|\tau_1) \psi_s(\mu_2 |\tau_2)\rangle=
  -\frac{\hbar
c}{\pi}\frac{1}{(\gamma (\tau_2 - \tau_1)c)^2- 4 r^2 \sin^2
\frac{\Omega \gamma (\tau_2 - \tau_1)}{2}}.
\end{eqnarray}
\indent This correlation function is also identical to the positive frequency Wightman function \cite{davis1996}(3), up to a constant. This function does not expose thermal features. Nevertheless the situation changes if the CF periodicity is taken into consideration. In the scalar field,  the CF can be considered
 periodical for the same reasons it is periodical in the electromagnetical fields. This issue is investigated below.\\
\subsection{Periodicity of the Correlation Function, Abel-Plana
Formula, and the Planck's Factor.}
\label{sec:PeriodictyScalaeCF}
 \indent  To take into consideration the periodicity of the CF we have to use its
  expression  ( \ref{eq:ScalarCF}) before integration over $\omega$. The equation
(\ref{eq:ScalarCF}) can be given in the form
\begin{eqnarray}
 \langle \psi_s(\mu_1 |\tau_1) \psi_s(\mu_2|\tau_2)\rangle=
 \frac{\hbar c k^2_0}{4 \pi ^2}\int
d O \int d \kappa \kappa \cos \kappa F_d, \;\; dO =\sin \theta d \theta d \phi, \;\;
\kappa =\frac{k}{k_0}, \;\; k_0 =\Omega/c, \;\; F_d = k_0 F.
\end{eqnarray}
If this function of $\tau = \tau_2 - \tau_1$ is  periodic then, as we saw for the CF
 $I^E_{(11)}$ above,  $\kappa =\frac{c k}{c k_0}=n=0,1,2,..$ and the
integral over $\kappa$ becomes an infinite series:
\begin{eqnarray}
\label{eq:ScalerCFd}
 \langle \psi_s(\mu_1 |\tau_1) \psi_s(\mu_2 |\tau_2)\rangle_d=
 \frac{\hbar c k^2_0}{4 \pi ^2}\int
d O \sum_{n=0}^{\infty} n \cos n F_d.
\end{eqnarray}
 Expression (\ref{eq:ScalerCFd} )is a definition of a new correlation function, with periodicity,  of the scalar massless field at the rotating detector. \\
 \indent The Abel-Plana summation formula
in this case is
\begin{eqnarray}
\sum_{n=0}^{\infty}n \cos nF_d =\int_0^{\infty} dt\; t \cos tF_d -
\int_0^{\infty} dt \frac{2t \cosh tF_d }{e^{2 \pi t}-1}
\end{eqnarray}
or
\begin{eqnarray}
\label{eq:planck} \Omega^2 \;\sum_{n=0}^{\infty}n \cos nF_d
=\int_0^{\infty} d\omega\; \omega \cos \omega \tilde{F} -
\int_0^{\infty} d\omega \frac{2\omega \cosh \omega \tilde{F}
}{e^{\frac{\hbar \omega}{k T_{rot}}}-1},
\end{eqnarray}
where $\tilde{F}=F_d / \Omega$ and $T_{rot}$ is defined in (\ref{eq:T}). \\
Then
\begin{eqnarray}
\langle \psi_s(\mu_1 |\tau_1) \psi_s(\mu_2 |\tau_2)\rangle_d=
 \frac{\hbar}{4 \pi ^2 c}\int d O \; \{ \;\int_0^{\infty} d\omega\; \omega \cos \omega \tilde{F} -
\int_0^{\infty} d\omega \frac{2\omega \cosh \omega \tilde{F}
}{e^{\frac{\hbar \omega}{k T_{rot}}}-1}\; \}
\end{eqnarray}
 \indent The  expression in $\{ \;\; \}$ is similar to the
right side of the  expression \cite{boyer1980}, (27) for the
correlation function of the scalar massless zero-point field at the detector at
rest in Planck's spectrum at the temperature T
\begin{eqnarray}
\int_0^{\infty}d \omega \omega \coth\frac{\hbar \omega}{2
kT}\cos\omega t =\int_0^{\infty}d \omega \cos \omega t +
\int_0^{\infty}d \omega \frac{2 \omega \cos\omega
t}{e^{\frac{\hbar \omega}{kT}}-1}.
\end{eqnarray}
 The appearance of the Planck
factor $(e^{\frac{\hbar \omega}{kT_{rot}}}-1)^{-1}$  shows
similarity between the radiation spectrum observed at the rotating detector
in the massless scalar zero-point field  and the radiation
spectrum observed by an inertial observer placed in a thermostat
filled with the radiation at the temperature $T=T_{rot}$. But
there is also a difference  between them. The $\tilde{F}$  and $\cosh$ are used in
the first expression whereas t and $\cos$ are used in the second
expression respectively.
 The $\tilde{F}$ is a function of $\theta$ and $\phi$. It means that a
thermal radiation observed by the rotating detector moving in the
massless scalar
zero-point radiation is anisotropic.  \\
 \indent  The  resemblance between both expressions  becomes closer
 if $t=0$ and
$\tilde{F}=0$ and two points of an observation agree. Both
expressions are identical. But in the case of one-point
observation which occurs when $\tilde{F}=0$ it is better to
consider the energy density of the scalar massless field, as is done
in the next section.
\subsection{The Energy Density and Planck's Spectrum.}
\label{sec:scalarEnergyDensity}
\indent The energy density $\langle
T_{(44)}\rangle$ of the massless scalar field at  the
detector rotating through the zero-point massless scalar field can be expressed in terms of the tensor of energy-momentum
$T_{ik}$ at the location of the detector in the laboratory
coordinate system \cite{synge} as
\begin{eqnarray}
\langle T_{(44)}\rangle=\mu_{(4)}^i\mu_{(4)}^k \langle
T_{ik}, \rangle
\end{eqnarray}
where $\mu_{a}^i$ are  tetrads. The  energy-momentum tensor is
\cite{birell1982}(2.27)
\begin{eqnarray}
T_{ik}= \psi_{,i}\psi_{,k}
-\frac{1}{2}\eta_{ik}\eta^{rs}\psi_{,r}\psi_{, s}, \;\; \eta_{ik}=
\eta_{ik}=diag(1,1,1,-1)
\end{eqnarray}
Using (\ref{eq:scalarField}), (\ref{eq:scalarTheta}), and
Frenet-Serret tetrads it is easy to show that
\begin{eqnarray}
\langle T_{11}\rangle=\langle T_{22}\rangle=\langle
T_{33}\rangle=\frac{1}{3}\langle T_{44}\rangle = \frac{\hbar
c}{3 \pi} \int dk k^3 =\frac{\hbar \Omega^4}{3 \pi c^3} \int d\kappa
\;\; \kappa^3
\end{eqnarray}
and
\begin{eqnarray}
\langle T_{(44)}\rangle = \frac{4 \gamma^2 -1}{3} \langle
T_{44} \rangle = \frac{4 \gamma^2 -1}{3}  \frac{\hbar
\Omega^4}{\pi c^3} \int d\kappa \;\; \kappa^3
\end{eqnarray}
  With periodical features taken
into consideration this expression has the following form
\begin{eqnarray}
\langle T_{(44)}\rangle_d  = \frac{4 \gamma^2
-1}{3}\;\;\frac{\hbar}{\pi c^3}\;  \Omega^4\; \sum_{n=0}^{\infty}
n^3
\end{eqnarray}
( It has an additional factor $n^2$ compared with (\ref{eq:ScalerCFd}) because
$T_{ik}$ have derivatives of $\psi$-functions. )
or
\begin{eqnarray}
\langle T_{(44)}\rangle_d =\frac{4 \gamma^2
-1}{3}\;\;\frac{\hbar}{\pi c^3}\;2 \;( \; \int _0^{\infty} d \,
\omega \frac{1}{2}\omega^3 +
 \int_0^{\infty} d\omega \frac{ \omega^3 }
 {e^{\hbar\omega/kT_{rot}} -1}.
\;)
\end{eqnarray}
\indent Let us compare this expression and the expression for the
energy density of the massless scalar field  with Planck's
spectrum of random thermal radiation at the temperature T,  along
with the zero-point radiation in an inertial reference frame,
\begin{eqnarray}
\langle T_{44} \rangle_T=\frac{1}{2} [(\frac{\partial
\psi_T}{\partial (ct)})^2 + (\frac{\partial \psi_T}{\partial x
})^2 + (\frac{\partial \psi_T}{\partial y})^2 + (\frac{\partial
\psi_T}{\partial z})^2],
\end{eqnarray}
where \cite{boyer1980}
\begin{eqnarray}
\psi_T =\int d^3k\; f_T (\omega)\; \cos\;[\;\vec{k}\vec{r} -
\omega t -\theta(\vec{k})\;]
\end{eqnarray}
and
\begin{eqnarray}
f^2_T(\omega)=\frac{c^2}{\pi^2}\; \frac{\hbar}{\omega}\; [\;
\frac{1}{2} + \frac{1}{\exp(\hbar \omega / k T) -1 }\;].
\end{eqnarray}
It is easy to show  that
\begin{eqnarray}
\langle T_{(44)} \rangle_d= \frac{2 (4 \gamma^2 -1)}{9}
\langle T_{44} \rangle_{T=T_{rot}}.
\end{eqnarray}
\indent So, due to periodicity of the motion, an observer rotating
through a zero point radiation of a massless random scalar field
should see the same energy density as  an inertial
observer would see, moving in a thermal bath at the temperature
$T_{rot}=\frac{\hbar \Omega}{2 \pi k}$,  multiplied by the
factor $\frac{2}{9}(4\gamma^2-1)$. This factor comes from
integration over angles and  is a consequence of anisotropy of the
scalar field measured by an observer with angular velocity $\Omega$.
\section{Conclusion and Perspectives.}
\label{sec:Discusssion}
\indent The thermal effects of non inertial motion investigated in
the past for uniform acceleration through classical random
zero-point radiation of electromagnetic and massless scalar field
are shown to exist in the case of
rotation motion as well. \\
\indent The rotating reference system $\{\mu_\tau\}$, along with
the two-point correlation  functions (CFs) and energy density, are
defined and used as the basis for investigating  effects observed
by  a detector rotating through random classical zero-point
radiation. The reference system consists of Frenet -Serret
orthogonal tetrads
 $\mu_\tau$. At each proper time $\tau$ the rotating detector is
 at rest  and has a constant acceleration vector at the
 $\mu_\tau$.
 \\
\indent The two-point CFs and the energy density at the rotating
reference system should be periodic with the period $T=\frac{2
\pi}{\Omega}$, where $\Omega$ is an angular detector velocity,
because CF and energy density measurements are one of the tools the
detector can use to justify the periodicity of its motion. The CFs
 have been calculated  for both \emph{electromagnetic} and
 \emph{massless scalar} fields in two cases, with and without taking
  this periodicity into
consideration. It was found that only periodic CFs have some
thermal features and  particularly the Planck factor with the
temperature $T_{rot}=\frac{\hbar \Omega}{2 \pi k_B}$ ($k_B$ is the
Boltzman constant). Mathematically this property is connected with
the discrete spectrum of the periodic CFs,   and its interpretation is
based on the Abel-Plana
summation formula. \\
\indent It is also shown that energy densities of the
electromagnetic and massless scalar fields observed by the
 detector rotating through classical zero-point radiation at zero
 temperature  are respectively
\begin{eqnarray} \nonumber
w = \frac{2 \;(4\gamma^2-1)}{3 }\; w_{em}(T_{rot})
\end{eqnarray}
and
\begin{eqnarray} \nonumber
\langle T_{(44)} \rangle_d= \frac{2 (4 \gamma^2 -1)}{9}
\langle T_{44} \rangle_{T_{rot}}.
\end{eqnarray}
Each of them consists of two terms. The first term, corresponding to zero-point radiation energy density, is divergent, and the second one, describing the thermal effect, is convergent.\\
\indent Let us discuss the convergent electromagnetic thermal energy density
\begin{eqnarray}
w_{em, T} = \frac{2 \;(4\gamma^2-1)}{3 } \times  \frac{4 \sigma}{c} T^4_{rot}, \nonumber \\
\gamma^2 = (1 - (\Omega r)^2 /c^2)^{-1}.
\end{eqnarray}
It includes factor  $\frac{2}{3}(4\gamma^2-1)$. Appearance of this
factor is connected with the fact that rotation is
defined by two parameters, angular velocity and the radius of
rotation, in contrast with a uniformly accelerated linear motion
which is defined by only one parameter, acceleration $a$. If, for a fixed $\Omega$, the radius of a circular orbit grows, $r \rightarrow c/ \Omega$, the second factor does not change but the first one grows.
Such behaviour of the convergent term may have a mechanical interpretation. \\
\indent Let several small particles with the same sign charge move
through the vacuum field on a circular orbit.
Let us further assume that repulsive interaction
of the particles  results in a shift of the particles  to
another circular orbit with slightly greater radius $r$ but with the same
angular velocity $\Omega$. Then the thermal energy  density $w_{em,T}$, observed locally by each of the particles, would increase. This increase demands an additional work against the vacuum field and therefore initiates  the force, let us call it $the \;\; vacuum \;\; force$,  which acts on these particles from the vacuum field. The volume density of this force is given by
\begin{eqnarray}
\label{eq-Force}
f_{vac} = - \frac{d w_{em,T}}{d r} = -\frac{8}{3} \frac{\Omega^2}{c^2}
\times \frac{2 r}{(1 - (\Omega r)^2/c^2)^2
  } \times \frac{4 \sigma}{c} T^4_{rot}
\end{eqnarray}
The force $f_{vac}$  does not depend on the size of neither the charge nor the mass and originates from the thermal energy $w_{em,T}$, even though it is positive.
These three features make $f_{vac}$ similar to the force, $f_{cas}$, in the Casimir  model for a charged particle \cite{Milonni1994, Boyer1968, Davis1972}
\begin{eqnarray}
E(a)= -C \frac{\hbar c}{2a}, \;\;\; f_{cas}= -\frac{d E}{da}=-C \frac{\hbar c}{2 a^2},
\end{eqnarray}
where $a$ is a radius.
This model was designed to explain a charged particle  stability. The force $f_{cas}$ also does not depend on the size of neither the charge nor the mass, the energy $E(a)$ is positive ( because $C \approx -0.09$ ) \cite{Boyer1968}. \\
\indent Nevertheless  $f_{vac}$ and $f_{cas}$ are significantly different. Indeed,   \\
1. The $f_{vac}$ is applied to the dynamical system of a particle (particles) moving on a circular orbit,  not to a static one as $f_{cas}$ in the Casimir model.\\
2. The $f_{vac}$ is attractive one  because it is directed from the location with greater positive energy density, $w_{em,T}$, to the location with smaller one, that is to the center of the circular orbit. Thus we could expect  it might balance the repulsive force associated with interaction of the charged particles. In contrast, the $f_{cas}$ is known to be repulsive, directed from the center of the shell outward, and therefore can not balance repulsive electrical forces.\\
3. The  $f_{vac}$ infinitely grows with $r \rightarrow \; r_0 = c/ \Omega$ and works as a restoring spring force. Therefore the radius of circular orbits with a fixed $\Omega$ is bounded. The orbits with a radius greater than $r_0$ do not
exist because the vacuum force becomes infinite. On the uttermost orbit with the radius
$r_0$, a linear velocity of the rotating particle would have become $c$\\
4. The  $f_{vac}$ becomes very small and proportional to $r$ when $r$ is small,
$r \ll c/\Omega$.\\
\indent The last two features of the $f_{vac}$ mean that the further the rotating particle from the center is the more bounded  it becomes or, in other words, confined. The closer to the center it is the freer it becomes. This reminds us of two significant concepts in quantum chromodynamics (QCD), the theory of strong interactions: asymptotic freedom and confinement. \\
\indent Confinement theory of quarks and gluons is still a challenge for strong interaction physics \cite{Gribov1999}. Therefore a concept of a newly introduced vacuum force can be useful for understanding a confinement phenomenon, even though the concept is introduced  in the frame of the stochastic electrodynamics but confinement realizes in strong interactions. Moreover quarks, with strong interaction between them, do have an electrical charge, can interact with the electromagnetic field vacuum and
experience the vacuum force.
In Appendix \ref{sec-Esimation} we make  rough and preliminary estimates of the
 $f_{vac}$ and $T_{rot}$, just to understand what order of magnitude they could have in hadron. \\
\indent This is only one of possible directions of the vacuum force $f_{vac}$ applications.
More detailed discussion of the vacuum force  will be given in  a different publication.\\
\indent The same consideration is true for the convergent thermal part of massless scalar field.   Some of the results discussed in this paper have been
obtained in \cite{levin2006}. \\
\section{ Acknowledgement. }
I am very thankful to prof. T.H. Boyer and prof. D.C. Cole for their encouraging
comments while reading my manuscript. In response to their recommendations  I have included the part about physical sense of the thermal properties in the section Conclusion and Perspectives.\\ \\
{\bf APPENDIX} \\
\appendix
\section{ Orthogonal Tetrads. }
\label{sec:Fermi-Walker}
\indent  An orthogonal tetrad ( OT ) is a  set of four orthogonal and normalized 4-vectors $\mu^i_{(a)}$,
labeled by a=1,2,3,4, so that
\begin{eqnarray}
\label{eq:OT1}
\mu^i_{(a)} \mu_{(b)i } = \eta_{(a b)}.
\end{eqnarray}
Co-vectors $\mu_i^{(b)}$ are defined as
\begin{eqnarray}
\label{eq:OT2}
\mu^{(a)i}= \eta^{(ab)} \mu^i_{(b)}, \;\;\; \mu^i_{(a)}=\eta_{(ab)} \mu^{(b)i}.
\end{eqnarray}
The $\eta_{(ab)}$ is a diagonal matrix
\begin{eqnarray}
\label{eq:OT3}
\eta_{(ab)}= \eta^{(ab)} = diag (1,1,1,-1).
\end{eqnarray}
 Frenet-Serret OTs satisfy the formulas \cite{synge}(55):
\begin{eqnarray}
\label{eq: Frenet-Serret_eqns}
D\mu^i_{(4)}=b\mu^i_{(1)}, \nonumber \\
D\mu^i_{(1)}= \tilde{c}\mu^i_{(2)} + b \mu^i_{(4)}, \nonumber \\
D\mu^i_{(2)}=d \mu^i_{(3)} -\tilde{c} \mu^i_{(1)}, \nonumber \\
D\mu^i_{(3)}=-d \mu^i_{(2)},
\end{eqnarray}
 where $D=\frac{d}{d \tau}$, $\tau$ is a proper time of the detector, in the flat space-time
with metric $g_{ik}$=diag(1,1,1,-1). Solution of this system is given in (\ref{eq:FStetrad}), with
\begin{eqnarray}
 b=-\beta \Omega \gamma^2, \; \tilde{c}=
 \Omega \gamma^2, ;\ d=0.
\end{eqnarray}
Fermi-Walker tetrad vectors  are defined as (\cite{moller72} (9.148, 4.139, and
4.167))
\begin{eqnarray}
\label{eq:FW}
\frac{d e_{(a)k}}{d \tau}= (e_{(a)l}\dot{U}_l)U_k/c^2
-(e_{(a)l}U_l)\dot{U}_k/c^2
\end{eqnarray}
and can be given in the form \cite{moller72} 4.167
\begin{eqnarray}
e_{(1)k}= (\cos\alpha \cos \alpha \gamma + \gamma \sin\alpha
\sin\alpha\gamma,\; \sin\alpha \cos \alpha \gamma - \gamma \cos
\alpha \sin \alpha \gamma,0,-i(v \gamma/c)\sin \alpha \gamma),
\nonumber \\
e_{(2)k}= (\cos\alpha \sin \alpha \gamma - \gamma \sin\alpha
\cos\alpha\gamma,\; \sin\alpha \sin \alpha \gamma + \gamma \cos
\alpha \cos \alpha \gamma,0,+ i(v \gamma/c)\cos \alpha \gamma),
\nonumber \\
e_{(3)k}=(0, 0, 1, 0), \nonumber \\
e_{(4)k}= (i \frac{v}{c}\gamma \sin \alpha, -i\frac{v}{c}\gamma
\cos \alpha ,0 ,\gamma).
\end{eqnarray}
In (\ref{eq:FW}) , as in \cite{moller72} ( 4.167 ),  the metric is chosen in the form
$g_{ik}=(1,1,1,1)$.\\
\indent We preferred to use Frenet-Serret tetrads, and  not Fermi-Walker ones, because
in  a reference frame associated with a Fermi-Walker tetrad  $e_{(a)i}$  the 3-vector
acceleration is not constant in both direction and magnitude
\begin{eqnarray}
\label{eq:FermiWalkerAcceleration} \dot{U}_{(a)}=
e_{(a)l}\dot{U}_l= (-a \Omega^2\gamma^2 \cos \alpha \gamma,-a
\Omega^2 \gamma^2 \sin \alpha \gamma,0,0),
\end{eqnarray}
and the acceleration depends on proper time $\tau$.
\section{ Some Correlation
Functions of an Electromagnetic Field at a Rotating Detector as 3-Dimensional Integrals over( $k, \theta, \phi $ ).} \label{sec-generalExpressions}
\indent Two correlation functions mentioned at the end of Section \ref{sec:TetradsMeasuments} are the following:
\begin{eqnarray}
I_{(22)}^E =\langle E_{(2)}(\mu_1 | \tau_1)\:E_{(2)}(\mu_2| \tau_2)\rangle=
\langle E_1(\tau_1)E_1(\tau_2)\rangle \sin \alpha_1 \sin \alpha_2
+ \langle E_1(\tau_1)E_2(\tau_2)\rangle
(-1)\sin \alpha_1 \cos \alpha_2 + \nonumber \\
\langle
E_2(\tau_1)E_1(\tau_2)\rangle (-1)\cos\alpha_1 \sin \alpha_2 +
\langle E_2(\tau_1)E_1(\tau_2)\rangle \cos \alpha_1 \cos \alpha_2, \nonumber \\
I_{(33)}^E = \langle E_{(3)}(\mu_1 | \tau_1)\:E_{(3)}(\mu_2| \tau_2)\rangle=
\gamma^2 \langle E_3(\tau_2)E_3(\tau_1) \rangle
-\gamma^2\frac{v}{c} \cos \alpha_1 \langle
E_3(\tau_2)H_1(\tau_1)\rangle - \gamma^2\frac{v}{c} \cos \alpha_2
\langle H_1(\tau_2)E_3(\tau_1)\rangle - \nonumber
\\
\gamma^2\frac{v}{c} \sin \alpha_1 \langle
E_3(\tau_2)H_2(\tau_1)\rangle-  \gamma^2\frac{v}{c} \sin \alpha_2
\langle H_2(\tau_2)E_3(\tau_1)\rangle+  \nonumber \\ \gamma^2(\frac{v}{c})^2
\cos \alpha_2 \cos \alpha_1  \langle H_1(\tau_2)H_1(\tau_1)\rangle
+
\gamma^2(\frac{v}{c})^2 \sin \alpha_2 \sin \alpha_1 \langle
H_2(\tau_2)H_2(\tau_1)\rangle + \nonumber \\   \gamma^2(\frac{v}{c})^2 \cos
\alpha_2 \sin \alpha_1 \langle H_1(\tau_2)H_2(\tau_1)\rangle +
\gamma^2(\frac{v}{c})^2 \sin \alpha_2 \cos \alpha_1 \langle
H_2(\tau_2)H_1(\tau_1)\rangle.
\end{eqnarray}
Is is easy to show that they  depend on the difference
$\delta=\alpha_2 -\alpha_1$ only. \\
\begin{eqnarray}
I^{E}_{(22)}=
\cos\delta \int d^3 k \;R + \sin^2 \frac{\delta}{2}\int d^3 k
\hat{k}^2_x \;R + (-1)\cos^2\frac{\delta}{2} \int d^3 k
\hat{k}^2_y \;R.
\end{eqnarray}
\begin{eqnarray}
I^E_{(33)}=
\gamma^2 \frac{v^2}{c^2} \cos \delta\;\int d^3 k\;R + \gamma^2
\frac{v}{c}(-2)\cos\frac{\delta}{2} \int d^3 k\;\hat{k}_y \;R+
\nonumber \\
\gamma^2[1-\frac{v^2}{c^2}\cos^2 \frac{\delta}{2}]\;\int d^3 k
\hat{k}^2_x \;R +
\gamma^2[1+\frac{v^2}{c^2}\sin^2\frac{\delta}{2}]\int d^3
k\;\hat{k}^2_y\;R.
\end{eqnarray}
Expressions for $R$ and $\delta$ are given in (\ref{eq-expressionF}). \\
\indent The non diagonal components of the correlation function
are zeroes :
\begin{eqnarray}
\langle E_{(1)}(\mu_1 | \tau_1)\:E_{(2)}(\mu_2| \tau_2)\rangle=
\langle E_{(1)}(\mu_2 | \tau_2)\:E_{(2)}(\mu_1| \tau_1)\rangle=0,
\nonumber \\
\langle E_{(1)}(\mu_1 | \tau_1)\:E_{(3)}(\mu_2| \tau_2)\rangle=
\langle E_{(1)}(\mu_2 | \tau_2)\:E_{(3)}(\mu_1| \tau_1)\rangle=0,
\nonumber \\
\langle E_{(2)}(\mu_1 | \tau_1)\:E_{(3)}(\mu_2| \tau_2)\rangle=
\langle E_{(2)}(\mu_2 | \tau_2)\:E_{(3)}(\mu_1| \tau_1)\rangle=0,
\end{eqnarray}
Similar expressions have been received for the CF with magnetic
field components. So all CFs can be given as 3-dimensional
integrals over $(k, \theta, \phi)$.
\section{Integral calculations:
final expression for $I_{(11)}^E $.}
\label{sec-IntegralCalculationFinal} \indent All non zero
expressions for the CF in subsection (\ref{sec-Final}) should be
integrated over $k, \theta$ , and $\phi$. The integral over $k$
can be easily calculated:
\begin{eqnarray}
\label{eq:IntegralOverK}
 \int_0^\infty d k k^3 \cos \{ k ( 2 r \sin
 \frac{\delta}{2} \sin \theta \sin \phi -c (t_2 -t_1))\} =  \frac
{6}{ {\{2 r \sin \frac{\delta}{2}  \sin \theta \sin \phi -
c(t_2-t_1)\}^4}}= \nonumber \\
= \frac {6}{[ c(t_2-t_1)]^4} \frac{1}{ [1-\frac{v}{c}\frac{\sin
\delta/2}{\delta/2}\sin\theta \sin \phi]^4 }.
\end{eqnarray}
The integrals over $\theta$ and $\phi$ can be represented  in
terms of elementary functions. Let us show it  for $I^E_{(11)} \equiv \langle
E_{(1)}(\mu_1|\tau_1)E_{(1)}(\mu_1|\tau_2) \rangle$:
\begin{eqnarray}
 I^E_{(11)} =
  \frac{3\hbar c}{2 \pi^2
[c(t_2-t_1)]^4} \gamma^2 \int_{0}^{\pi} d\theta
\times \{ \;\; (\cos \delta \sin \theta +(-\cos^2
\frac{\delta}{2}+ \frac{v^2}{c^2}) \sin^3 \theta) \int_{0}^{2\pi}
d\phi \:
\frac{1}{(1+b \sin\phi)^4} \nonumber \\
+ (-2\frac{v}{c}\cos\frac{\delta}{2})sin^2\theta\int_{0}^{2\pi}
d\phi \: \frac{\sin\phi}{(1+b \sin\phi)^4}
+\sin^3\theta\int_{0}^{2\pi} d\phi \: \frac{\sin^2\phi}{(1+b
\sin\phi)^4} \;\; \},
\end{eqnarray}
We have taken into consideration here that
\begin{eqnarray}
\label{eq:HatVector} \hat{k}_x = \sin\theta \cos \phi ,  &
\hat{k}_y= \sin\theta \sin\phi ,  & \hat{k}_z= \cos \theta
\end{eqnarray}
and used notations  $b \equiv k \: \sin\theta, \;\; k \equiv
-\frac{v}{c}\frac{\sin\delta/2}{\delta/2}$. So k is a constant,
not a wave vector.  \\
 \indent The next step is to calculate the integral over
$\phi$. Because \cite{gr1965},
\begin{eqnarray}
\int_0^{2\pi}d\phi \frac{1}{(1+b \sin \phi)^4}=\frac{\pi(2 + 3
b^2)}{(1-b^2)^{7/2}},
\end{eqnarray}
\begin{eqnarray}
\int_0^{2\pi}d\phi \frac{\sin \phi}{(1+b \sin \phi)^4}=
\frac{-b\pi (4+b^2)}{(1-b^2)^{7/2}},
\end{eqnarray}
and
\begin{eqnarray}
\int_0^{2\pi}d\phi \frac{\sin^2 \phi}{(1+b \sin \phi)^4}=
\frac{\pi(1+4b^2)}{(1-b^2)^{7/2}},
\end{eqnarray}
the correlation function takes the form (\ref{eq-CF11Final}).
\section{Another Way to Receive the $I^E_{11)}$ for the Discrete Spectrum. }
\label{sec-ModifiedExpressionRandom}  In  section
(\ref{sec:CFDiscrete}) we have obtained the general expression for
the CF $I^E_{(11)d}\equiv \langle E_{(1)}(\mu_1|\tau_1)
E_{(1)}(\mu_2|\tau_2)\rangle_d$ with discrete spectrum, based on its periodicity.
This also could be done directly using the following expressions for the fields $E_i$ and $H_i$, with discrete spectrum, instead of the equations (\ref{eq:ff}):
\begin{eqnarray}
\vec{E}(\vec{r},t)= a\:\sum^{\infty}_{n=0} \sum^2_{\lambda=1} \int
do\,k^2_n\,\hat{\epsilon}(\hat{k},\lambda)\,h_0(\omega_n)\
\cos[\vec{k_n}\vec{r}-\omega_n
t -\Theta(\vec{k_n},\lambda)], \nonumber \\
\vec{H}(\vec{r},t)=a\:\sum^{\infty}_{n=0}\sum^2_{\lambda=1} \int
do \,k^2_n \,
[\hat{k},\hat{\epsilon}(\hat{k},\lambda)]\,h_0(\omega_n)\,\cos[\vec{k_n}\vec{r}-\omega_n
t-\Theta(\vec{k_n},\lambda)], \nonumber \\
 \vec{k}_n=k_n \hat{k},
\;\; k_n=k_0\,n, \;\; k_0= \frac{\Omega}{c}, \;\; \omega_n=c
\,k_n, \;\;
 do= d\theta \, d\phi \, \sin\theta, \nonumber \\
 \hat{k}=(\hat{k}_x,
\hat{k}_y, \hat{k}_z)=(\sin\theta \, \cos \phi, \, \sin\theta \,
\sin\phi, \,\cos\theta\,), \;\; a = k_0. \label{eq:mff}
\end{eqnarray}
 The unit  vector $\hat{k}$ defines a
direction of the wave vector $\vec{k}$ in a spherical momentum 3-space and does not depend on its value,
n.\\
The right side of the first equation in the relation
(\ref{eq:thetas}) should be modified. We do this in two steps. First
we rewrite them in a spherical momentum space  \cite{davydov1968},
p.656 as :
\begin{eqnarray}
\label{eq:cosaverage} \langle \cos\theta(\vec{k}_1 \lambda_1
)\cos\theta(\vec{k}_2 \lambda_2 ) \rangle=\langle
\sin\theta(\vec{k}_1 \lambda_1 )\sin\theta(\vec{k}_2 \lambda_2 )
\rangle=\frac{1}{2}\delta_{\lambda_1
\lambda_2}\delta^3(\vec{k}_1-\vec{k}_2)=
\frac{1}{2}\delta_{\lambda_1 \, \lambda_2} \, \frac{2}{k_1^2}\,
\delta(k_1-k_2)\delta(\hat{k}_1-\hat{k}_2).
\end{eqnarray}
And then, in the case of the discrete spectrum, it will be the following:
\begin{eqnarray}
\label{eq:cosaverage} \langle \cos\theta(\vec{k}_{n_1} \lambda_1
)\cos\theta(\vec{k}_{n_2} \lambda_2 ) \rangle= \langle
\sin\theta(\vec{k}_{n_1} \lambda_1 )\sin\theta(\vec{k}_{n_2}
\lambda_2 )\rangle=  \frac{1}{2}\delta_{\lambda_1 \, \lambda_2} \,
\frac{2}{k_0(k_0 n_1)^2}\, \delta_{n_1 \,
n_2}\delta(\hat{k}_1-\hat{k}_2).
\end{eqnarray}
The equation
 $\;\; \sum^2_{\lambda=1}\epsilon_i(\vec{k}\lambda
)\epsilon_j(\vec{k}\lambda )= \delta_{ij}-\hat{k}_i\hat{k}_j \;\;$
does not depend on n. The correlation function finally takes the
form (\ref{eq:CF11d1},\ref{eq:CF11d2})
\section{Expression for $S_d $ after Integration over $\omega$.}
\label{sec-anotherExpressionSd}
The  expressions for $S_d$ in ( \ref{eq:sum}) mentioned in subsection \ref{sec:Abel-Plana} ,  after integration over $\omega$,
are the following:
 \begin{eqnarray}
 \label{eq:I_one}
 S_d = \frac{6}{F_d^4} + [\frac{3 -2 \sin^2(F_d/2)}{8
 \sin^4(F_d/2)} -\frac{6}{F_d^4}]
 \end{eqnarray}
 or
\begin{eqnarray}
\label{eq:I_two}
 S_d= \frac{6}{F_d^4} +
6  \sum_{n=1}^{\infty} \frac{1}{(2\pi n)^4}\;\; [\;\; \frac{1}
{(1+ F_d /2\pi n)^4}
  + \frac{1}{(1- F_d /2\pi n)^4} \;\; ].
\end{eqnarray}
 The first integral of $S_d$  in ( \ref{eq:sum}) is  divergent and the second one
is convergent as  can seen from (\ref{eq:I_one}).
\section{Correlation Function Calculation for Random Zero-Point Radiation of a Scalar Massless Field }.
\label{sec:CF_Scalar_Field}
The expression (\ref{eq:ScalarCF}) after integration over positive $k= \omega /c$ becomes:
\begin{eqnarray}
\langle \psi_s(\mu_1 |\tau_1) \psi_s(\mu_2
|\tau_2)\rangle=-\frac{\hbar c}{4\pi^2}\:\int_0^\pi d\theta
\:\sin\theta \: \int_0^{2 \pi} d\phi \: \: [E \:\sin \phi-
B]^{-2},
\end{eqnarray}
where $B=\gamma \tau c$, $E=2 a \sin\theta \: \sin \frac{ \Omega
\gamma \tau}{2}$, and $\tau=\tau_2- \tau_1$. \\
Because $B-|E|= c \gamma \tau \{1-\frac{v}{c} |\sin \theta \:
\frac{\sin \pi (\gamma \tau /T)}{\pi(\gamma \tau / T)}|\}> c
\gamma \tau (1 - v/c) >0,$
 and using \cite{prudnikov} we obtain :
\begin{eqnarray}
\int_0^{2 \pi} d\phi \: \frac {1}{[E \:\sin \phi- B]^2}= \frac{2
\pi B}{(B^2 - E^2)^{3/2}}.
\end{eqnarray}
Finally, integrating it over $\theta$ we come to (\ref{eq:WightmanFunction}).
\section{The Force $f_{vac}$ and Temperature $T_{rot}$ Estimations.}
\label{sec-Esimation}
The only purpose of the following estimations is to figure out the order of a magnitude
of the vacuum force $f_{vac}$ and rotation temperature $T_{rot}$ which can be associted with a proton size $r_0 \approx 10^{-15}\; m$  \cite{Karshenboim2008}. \\
\indent The radial component of the force acting on a spherical particle of the radius $a$  rotating through  an
 electromagnetic zero-point field on a circular orbit
with radius $r\approx r_0$ and with angular velocity $\Omega= c/r_0$ can be given in
the form 
\begin{eqnarray}
F=f_{vac} \;\frac{4}{3}\pi a^3 =- \frac{x}{(1-x^2)^2}\; \frac{4 c \hbar}{135 \; \pi} \; \frac{a^3}{r_0^5}, \;\;\; x=\frac{r}{r_0} \leq 1,
\end{eqnarray}
where $f_{vac}$ is given in (\ref{eq-Force}).
Just for estimation purposes, let us take $a \approx 10^{-18} m $. Then
\begin{eqnarray}
F \approx -\frac{x}{(1-x^2)^2} \times (2.8)\times 10^{-7} \frac{J}{m}=  -\frac{x}{(1-x^2)^2} \times (1.75)\times 10^{-12} \frac{GeV}{Fermi}.
\end{eqnarray}
For $1-x \approx 10^{-6}$, $F \approx -(4.4)\times 10^{-1} \; GeV /Fermi = .7 \times
10^5 $ newtons  in a good agreement with an order of magnitude of strong interaction forces.\\
\indent Similarly,
\begin{eqnarray}
T_{rot}=\frac{\hbar \Omega}{2 \pi k_B} =\frac{\hbar c}{2 \pi k_B} \;\frac{1}{r_0},
\end{eqnarray}
and, for  distances  $r_0 \approx 10^{-15}m $, corresponds to the temperature $T_{rot} \approx 3.4 \times 10^{11} K $, a little bit less then the temperature $(1.90 \pm 0.02) \times 10^{12} K $ needed for  a quark-gluon plasma creation \cite{Fodor2004}.\\
\indent These estimations is a good motivation for further investigations.

\end{document}